# Rigorous Diffraction Interface Theory

Christopher M. Roberts,[1)] and Viktor A Podolskiy[1)]

[1]Department of Physics and Applied Physics, University of Massachusetts – Lowell, Lowell, Massachusetts, 01854, USA

We present a new formalism for understanding the optical properties of metasurfaces, optically thin composite diffractive devices. The proposed technique, Rigorous Diffraction Interface Theory (R-DIT), provides an analytical framework for understanding the transition between optically thin and optically thick structures. For metasurfaces, R-DIT avoids the calculation of optical propagation through thin layer and provides a direct link between the composition and geometry of a metasurface and its transmission, reflection, and diffraction properties.

Metasurfaces, optically thin structures with engineered diffraction, are emerging as a new ultra-compact platform for manipulating the flow of light. Applications of metasurfaces range from polarization conversion, to beam shaping, phase control, holography, and cloaking [1–13]. In contrast to bulk photonic devices whose optical response can be linked to dispersion of their bulk modes, the optics of metasurfaces is dominated by optically thin interfaces. As result, traditional analytical modeling methods, created with bulk materials in mind, are inefficient when applied to designing, understanding, and optimizing metasurfaces. Several techniques which reduce the optics of metasurfaces to generalized boundary conditions have been proposed[14–19], dramatically speeding up computations. However, the computational speedup is often accompanied by a loss of precision, especially for resonant or relatively thick (although still subwavelength) structures. The new technique presented in this work, Rigorous Diffractive Interface Theory (R-DIT), addresses the accuracy problem, while preserving the speed improvement of generalized boundary condition-based techniques. More importantly, R-DIT provides a missing link between bulk-based and surface-based photonics, playing a role of a Taylor expansion of the optical properties of complex composites in terms of their optical thickness.

The optical response of inhomogeneous composites, such as those shown schematically in Figure 1, are often analyzed with the help of numerical solutions of Maxwell's equations. The Finite Element Method (FEM)[20] and the Finite Difference Time Domain (FDTD)[21] method are considered to be state-of-the-art all-purpose electromagnetic solvers. However, these techniques are generally inefficient when multiple length scales are involved since they require meshing a relatively large region of space (often, a wavelength-scale unit cell) with elements that are substantially smaller than the smallest length-scale involved in the problem. As result, solutions are often time-, memory-, and CPU-intensive, to the point that renders optimization problems virtually intractable.

Mode-matching techniques, such as rigorous coupled wave analysis (RCWA), Fast Fourier Method (FFM), and others [22–24] offer significant memory efficiencies when analyzing the response of structures with relatively few interfaces and relatively thick (although, composite) layers. These techniques employ a two-step solution by representing the fields in each layers as a linear combination of well-behaved "modes". The first (often, most time-consuming) step lies in calculating the field profiles and the propagation properties of the modes, while the second step provides a relationship between the amplitudes of these modes throughout the system.

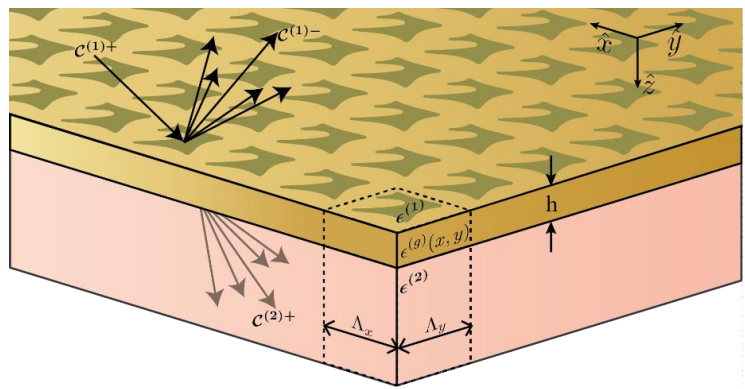

Figure 1: Schematic geometry of a metasurface like structure. R-DIT is applicable to structures where $h \ll \lambda$

Analytically, modal analysis is typically mapped onto an eigenvalue problem where eigenvalues represent the propagating constants of the modes and eigenvectors represent field profiles. In periodic media, shown schematically in Figure 1, application of Bloch theorem allows to represent a spatial profile of the individual mode inside the unit cell of the composite as a linear combination of Fourier components, with amplitudes of these components represented by the [vectors] of amplitudes $\vec{\mathcal{E}}, \vec{\mathcal{H}}$ (that in general depend on $z$ coordinate). Explicit substitution of such expansion into Maxwell equations, reduces them to

$$\frac{d}{dz}\begin{bmatrix}\vec{\mathcal{E}}\\\vec{\mathcal{H}}\end{bmatrix} = \frac{ic}{\omega}\widehat{\mathcal{M}}\begin{bmatrix}\vec{\mathcal{E}}\\\vec{\mathcal{H}}\end{bmatrix} \tag{1}$$

with elements of the matrix $\widehat{\mathcal{M}}$ relating the Bloch vectors, along with Fourier transforms of spatial profiles of material permittivity and permeability (see the Appendix and Refs.[22–24] for details). Assumption of harmonic propagation $[\vec{\mathcal{E}}, \vec{\mathcal{H}} \propto \exp(i\,k_z z)]$ transforms Eq.(1) into an eigenvalue problem,

$$\left(\frac{c}{\omega}\widehat{\mathcal{M}} - k_z\,\hat{I}\right)\begin{bmatrix}\vec{\mathcal{E}}\\\vec{\mathcal{H}}\end{bmatrix} = 0 \tag{2}$$

mentioned above. For homogeneous isotropic layers, solution to Eq.(2) yields a set of plane waves with $s$- and $p$-polarizations, often used in transfer-matrix or related formalisms [25].

However, the two-step mode-matching approach to Maxwell equations was again developed with bulk materials in mind. Metasurfaces often possess small optical thickness, thus, at least in principle, propagation of the field through metasurface can be neglected and, in the same limit, mode analysis of metasurface layer can be avoided. Several techniques have aimed to address this goal [14–18] and to directly relate the amplitudes of the modes in the layers surrounding the metasurface to each other

$$\left[\vec{\mathcal{C}}^{(l+1)}\right] = \widehat{\mathbb{T}}^{(l+1,l)}\left[\vec{\mathcal{C}}^{(l)}\right] \tag{3}$$

Each of these simplified techniques has its limitations. Thus, [18] focus on calculation of the main reflected/transmitted beam and neglect diffraction, a phenomenon that often dominates optical response of metasurfaces. The approach presented in [15] requires knowledge of the primary reflection and transmission to deduce the amplitudes of the diffracted beams. Diffractive interface theory (DIT) [14,19] provides the full solution of amplitudes of reflected and transmitted beams in very thin metasurfaces. In practice, the validity of DIT is limited to structures that are thinner than $1/10\ldots1/20$-th of the free space wavelength. Clearly, a better technique that would be capable of estimating the propagation of the modes through metasurface without solving Eq.(2) would provide a powerful tool to the photonics community. More importantly, such a technique would provide important insight into the emergence of propagating modes in the composites and allow understanding of the transition between surface- and bulk-dominated photonics. Below we present the formalism for developing such a technique.

We assume the time-harmonic dependence of the fields, $\vec{E}, \vec{H} \propto e^{-i\omega t}$, and follow the approach previously utilized to develop both RCWA[22,23] and DIT[14] formalisms. We first focus on the metasurface layer that we assume to be homogeneous along the $z$ coordinate and, starting from Maxwell equations in the component form, arrive to the four equations relating the in-plane components of electric and magnetic fields. For simplicity, we present results for the case when the components of metasurface have isotropic permittivity and permeability; generalization to anisotropic components is relatively straightforward, although cumbersome. Explicitly, Maxwell equations reduce to

$$\frac{d}{dz}\begin{bmatrix}E_x\\E_y\end{bmatrix} = \frac{ic}{\omega}\begin{bmatrix}-\partial_x\epsilon_g^{-1}\partial_y & \partial_x\epsilon_g^{-1}\partial_x + \mu_g\omega^2/c^2\\ -\partial_y\epsilon_g^{-1}\partial_y - \mu_g\omega^2/c^2 & \partial_y\epsilon_g^{-1}\partial_x\end{bmatrix}\begin{bmatrix}H_x\\H_y\end{bmatrix} \tag{4a}$$

$$\frac{d}{dz}\begin{bmatrix}H_x\\H_y\end{bmatrix} = \frac{ic}{\omega}\begin{bmatrix}\partial_x\mu_g^{-1}\partial_y & -\partial_x\mu_g^{-1}\partial_x - \epsilon_g\omega^2/c^2\\ \partial_y\mu_g^{-1}\partial_y + \epsilon\omega^2/c^2 & -\partial_y\mu_g^{-1}\partial_x\end{bmatrix}\begin{bmatrix}E_x\\E_y\end{bmatrix} \tag{4b}$$

For a wide class of metasurfaces with periodic in-plane profile, the fields inside a metasurface can be represented as linear combination of Bloch modes, $\{E, H\}_\alpha(x, y, z) = \sum_m \{\mathcal{E}(z), \mathcal{H}(z)\}_{\alpha,m} \exp i(q_{x,m}x + q_{y,m}y)$, where $\alpha$ represents Cartesian component, $\mathcal{E}$ and $\mathcal{H}$ stand for modal amplitudes of components of electric and magnetic fields, and the components of the pseudo-wavenumber $q$ are (properly arranged) multiples of reciprocal lattice (see, appendix and Ref[14] for details).

Substitution of the Bloch expansion into Eqs.(4) allows us to transform the differential Eq.(1) into a set of two first-order linear ODEs (that collectively describe Eq.(1) for metasurfaces with isotropic components):

$$\frac{d}{dz}\vec{\mathcal{E}}(z) = \frac{i\omega}{c}\hat{\mathbb{P}}\,\vec{\mathcal{H}}(z) \tag{5a}$$

$$\frac{d}{dz}\vec{\mathcal{H}}(z) = \frac{i\omega}{c}\hat{\mathbb{Q}}\,\vec{\mathcal{E}}(z) \tag{5b}$$

where components of the vectors $\mathcal{E}$ and $\mathcal{H}$ span indices $\alpha$ and $m$, described above. Note that geometry of unit cell, as well as spatial profile of permittivity and permeability has to be taken into account when deriving components of the matrices $\hat{\mathbb{P}}$ and $\hat{\mathbb{Q}}$ to maximize the stability of truncated Fourier expansion [26].

Eqs. (5a,5b) describe the evolution of the electromagnetic fields across the composite layer. For optically thick layers, Eqs.(5a,5b) yield an eigenvalue problem that has been described above[22–24]. Alternatively, Eqs.(5a,5b) can be used to *numerically* calculate the field across the metasurface using, e.g. trapezoidal integration[27,28]. For optically thin structures, however, Eqs.(5a,5b) can be used to *analytically* relate the fields across the metasurface using Taylor expansion:

$$\vec{\mathcal{E}}(z+\delta) \simeq \vec{\mathcal{E}}(z) + \left.\frac{d\vec{\mathcal{E}}}{dz}\right|_z \delta + \frac{1}{2}\left.\frac{d^2\vec{\mathcal{E}}}{dz^2}\right|_z \delta^2 + \cdots$$

$$= \vec{\mathcal{E}}(z) + \frac{i\delta\omega}{c}\hat{\mathbb{P}}\,\vec{\mathcal{H}}(z) - \frac{\delta^2\omega^2}{2c^2}\hat{\mathbb{P}}\hat{\mathbb{Q}}\,\vec{\mathcal{E}}(z) + \cdots \tag{6a}$$

$$\vec{\mathcal{H}}(z+\delta) \simeq \vec{\mathcal{H}}(z) + \left.\frac{d\vec{\mathcal{H}}}{dz}\right|_z \delta + \frac{1}{2}\left.\frac{d^2\vec{\mathcal{H}}}{dz^2}\right|_z \delta^2 + \cdots$$

$$= \vec{\mathcal{H}}(z) + \frac{i\delta\omega}{c}\hat{\mathbb{Q}}\,\vec{\mathcal{E}}(z) - \frac{\delta^2\omega^2}{2c^2}\hat{\mathbb{Q}}\hat{\mathbb{P}}\,\vec{\mathcal{H}}(z) + \cdots \tag{6b}$$

Eqs.(6a,6b) that represent the first main result of this work, provide a relationship between the fields across the metasurface and can be used to understand the emergence of field propagation and attenuation caused by non-zero thickness of the metasurface. Higher-order terms are necessary for analysis of electromagnetism in optically-dense composites (such as structures that have metallic, high-index, or high-loss inclusions).

Alternatively, Eqs.(6a,6b) can be used to directly relate the amplitudes of the modes in the layers that surround the metasurface, assuming that relationship between the fields and the modal amplitudes in these layers is known. For configuration shown in Figs. 2-5, such relationship can be written in the form of the matrix multiplication,

$$\begin{bmatrix}\vec{\mathcal{E}}(z)\\ \vec{\mathcal{H}}(z)\end{bmatrix} = \widehat{\mathbb{F}^{(l)}}(z)\begin{bmatrix}\vec{C}^{(l)+}\\ \vec{C}^{(l)-}\end{bmatrix} \tag{7}$$

The elements of matrix $\mathbb{F}$ for homogeneous layers are detailed in Ref.[14] and are provided in the appendix. Once the relationship between the modal amplitudes and the components of the Bloch expansion are known, the continuity of the in-plane fields across the metasurface interfaces along with Eqs.(6) can be used to relate the field at $z = \pm h/2$ and at $z = 0$, leading to the second main result of this work,

$$\begin{bmatrix} \mathbb{I} - \frac{h^2\omega^2}{8c^2}\hat{\mathbb{P}}\hat{\mathbb{Q}} + \cdots & i\frac{h\omega}{2c}\hat{\mathbb{P}} - \frac{ih^3\omega^3}{48c^3}\hat{\mathbb{P}}\hat{\mathbb{Q}}\hat{\mathbb{P}}\cdots \\ i\frac{h\omega}{2c}\hat{\mathbb{Q}} - \frac{ih^3\omega^3}{48c^3}\hat{\mathbb{Q}}\hat{\mathbb{P}}\hat{\mathbb{Q}}\cdots & \mathbb{I} - \frac{h^2\omega^2}{8c^2}\hat{\mathbb{Q}}\hat{\mathbb{P}} + \cdots \end{bmatrix}\widehat{\mathbb{F}^{(1)}}\left(-\frac{h}{2}\right)\begin{bmatrix}\vec{C}^{(1)+}\\ \vec{C}^{(1)-}\end{bmatrix}$$

$$= \begin{bmatrix} \mathbb{I} - \frac{h^2\omega^2}{8c^2}\hat{\mathbb{P}}\hat{\mathbb{Q}} + \cdots & -i\frac{h\omega}{2c}\hat{\mathbb{P}} + \frac{ih^3\omega^3}{48c^3}\hat{\mathbb{P}}\hat{\mathbb{Q}}\hat{\mathbb{P}}\cdots \\ -i\frac{h\omega}{2c}\hat{\mathbb{Q}} + \frac{ih^3\omega^3}{48c^3}\hat{\mathbb{Q}}\hat{\mathbb{P}}\hat{\mathbb{Q}}\cdots & \mathbb{I} - \frac{h^2\omega^2}{8c^2}\hat{\mathbb{Q}}\hat{\mathbb{P}} + \cdots \end{bmatrix}\widehat{\mathbb{F}^{(2)}}\left(\frac{h}{2}\right)\begin{bmatrix}\vec{C}^{(2)+}\\ \vec{C}^{(2)-}\end{bmatrix} \tag{8}$$

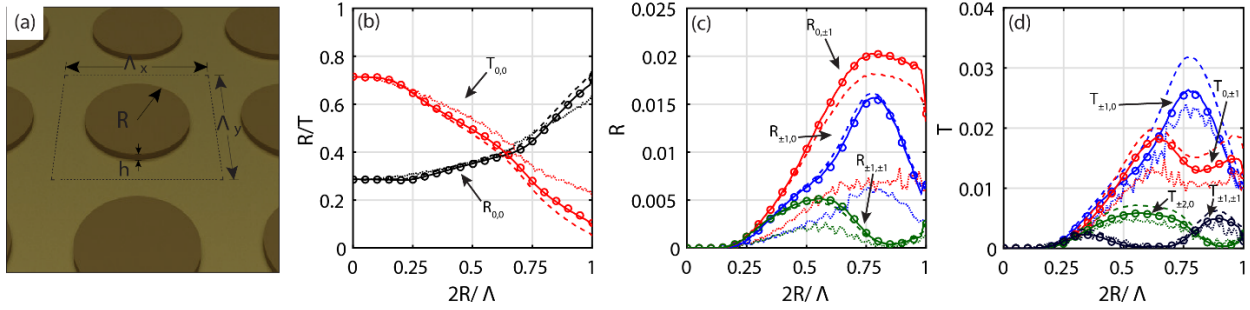

Figure 2: Diffraction by a metasurface formed by an array of plasmonic disks[$\epsilon_g = -10 + 1i, h = \lambda_0/10$] deposited on dielectric substrate [$\varepsilon_2 = 10.89$]; (a) schematic of the metasurface configuration; (b) 0th order reflection and transmission; (c),(d)higher-order reflection and transmission; open circles represent RCWA, dotted lines are the (original) DIT calculations, dashed and solid lines represent 1$^{st}$ and 3$^{rd}$ order RDIT respectively.

Generalized boundary condition, represented by Eq.(8), can be used to directly calculate reflection, transmission, and diffraction efficiency of the metasurface. Its leading (thickness-independent) term reduces Eq.(8) to conventional transfer-matrix formalism that neglects diffraction effects. The next, linear in $\delta$ term, contains Diffractive Interface Theory, originally presented in Ref.[14]. Higher order corrections rigorously extend the DIT to composites with increasingly higher optical thickness.

Note that increased precision of the generalized boundary conditions comes at the expense of additional matrix multiplications. Therefore, high precision implementation of R-DIT will approach the performance of iterative eigenvalue solvers, such as RCWA.

To illustrate the performance of R-DIT and compare its precision to other existing techniques, we analyze reflection properties of three metasurface structures, an array of plasmonic disks deposited on dielectric substrate, a free-standing checkerboard pattern, and polarization converter (originally reported in Ref.[1]), with original implementation of DIT[14], with R-DIT of different precision, as well as with RCWA.

We first consider the optical response of a periodic array ($\Lambda = 15.92 \mu m$) of plasmonic disks ($\epsilon = -10 + i$, realized, e.g. in the semiconductor-based "designer metal" paradigm[29]), deposited on GaAs substrate ($\epsilon_2 = 10.89$), operating at the free-space wavelength $\lambda_0 = 8 \mu m$. Figure 2 shows reflection and transmission for the composite as a function of disk radius, calculated using RCWA, DIT, and R-DIT for all diffraction orders for disk thickness of $\lambda_0/10$. It is seen that while DIT provides rough approximation of the optical response of this composite, R-DIT improves the accuracy of this approximation.

Figure 3 illustrates the utility of R-DIT to describe optical response of resonant metasurfaces on the example of polarization converter, originally proposed in Ref.[1]. Once again, it is seen that higher-order R-DIT is increasingly accurate in predicting the optical response of complex diffractive composites.

R-DIT can dramatically speedup the calculations of the optical response of thin diffractive composites embedded in layered optical structure. When implemented on a multi-threaded mulit-CPU system (4 core Xeon L5520 x 2 NUMA nodes)[30], first-order R-DIT is ~100% faster than an RCWA analysis; in practice a larger (but comparable) speed increase can be observed with CPU computing platforms.

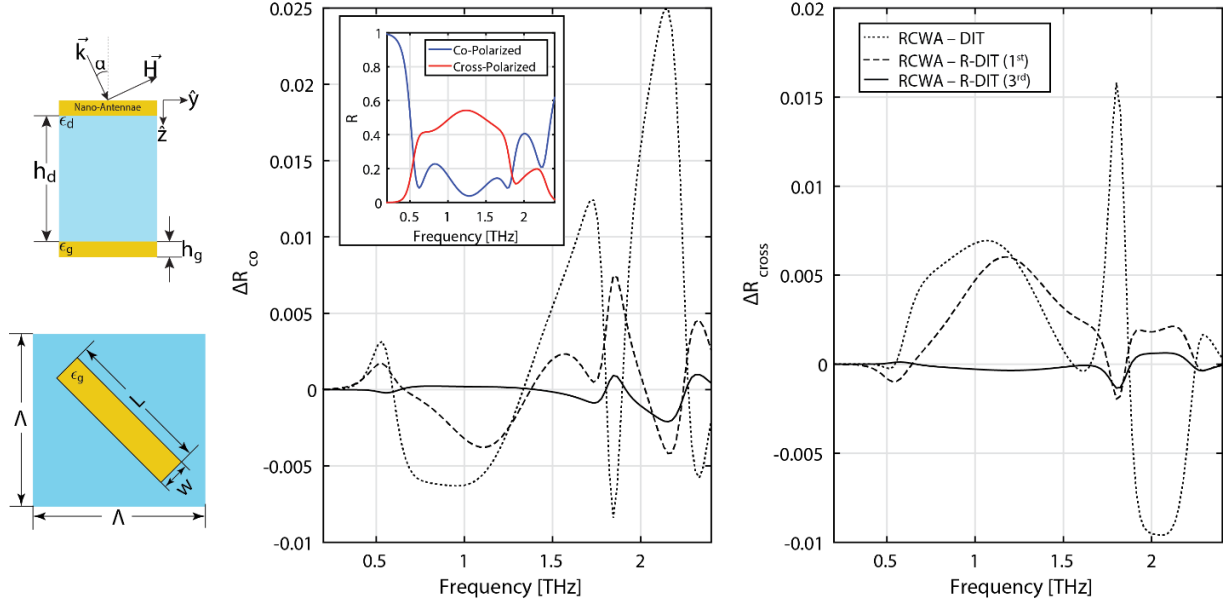

Figure 3: Optical properties of a nano-antennae polarization converter; (a) cross section of structure; (b) metasurface geometry; nano-antennae layer and gold ground layer are both 200nm thick (hg); periodicity $\Lambda_x = \Lambda_y = 68\mu m$, antenna wire length $L = 82\mu m$, antenna width $w = 10\mu m$, dielectic layer height $h_d = 33\mu m$ with permittivity of $\epsilon_d = 3(1 + 0.05i)$ and angles $\theta = 45°$, and $\alpha = 25°$. (c insert) Reflection of the metasurface results for co-polarized (blue) and cross-polarized (red) reflection calculated by RCWA. (c & d) Difference in co- & cross- polarized reflection between RCWA and DIT (dotted line), 1$^{st}$ order R-DIT (solid line) and 3$^{rd}$ order R-DIT (dashed)

A further, order of magnitude, speed increase can be achieved when R-DIT is implemented on (GP)GPU [31] hardware (Tesla K20c) that is specifically optimized for LU factorization-based algorithms (eigenvalue solvers, the cornerstone of RCWA do not enjoy similar GPU optimization at this time[32]). For example, on our workstation CPU-implemented R-DIT calculated one data point from Fig.2 in 4.1s; same calculations with RCWA calculations took 9.7s; GPU-implemented algorithms ran in 0.49s and 3.9s, respectively. We were unable to solve for optical properties of polarization converter with commercial direct-solver FEM[33] on our workstation. Given $\mathcal{O}(N^3)$ scaling typical of direct-solver FEM calculations, we estimate that these calculations would require ~500Gb RAM and would converge in ~24 hours to calculate a single data point.

To test the limits of R-DIT applicability, we calculated optical properties of the free-standing periodic ($\Lambda = 11.26\mu m$) vacuum-lossy dielectric ($\epsilon = 10 + i$) checkerboard pattern, operating at $\lambda_0 = 8\mu m$ as a function of the composite thickness. The checkerboard pattern is chosen as a "toy geometry" known for slow convergence of its optical properties with numerical solvers[24, 26]. It is seen that all implementations of DIT provide accurate description for relatively thin ($h < \lambda_0/20$) structures. 3$^{rd}$ order R-DIT improves the range of applicability of generalized boundary conditions up to $\lambda_0/10$-thick systems, while 10$^{th}$ order is applicable up to $\lambda_0/2$-thick systems (Figure 4). Importantly, it is seen that the higher order R-DIT expansions converge to the values obtained by full-wave RCWA. Note that even high-order R-DIT is still significantly faster than RCWA when implemented on GPU platform.

The deviation between R-DIT and RCWA (or, alternatively, convergence of multiple orders of R-DIT) can be used to understand the nature of optical response of the composite. Thus, when the optical properties of the system are described by 0-order ($h$-independent) terms, the contribution of the surface is negligible. Linear (in $h$) response indicates that the optical behavior is fully dominated by the interface, while the dynamics of bulk modes of the composite is not important. The importance of the 3$^{rd}$ (and higher) orders of R-DIT can be considered to be a "practical test" of emergence of light *propagation through* the structure. Note that the critical thickness at which the bulk properties emerge depends on both composition (permittivity) as well as on the geometry of composite.

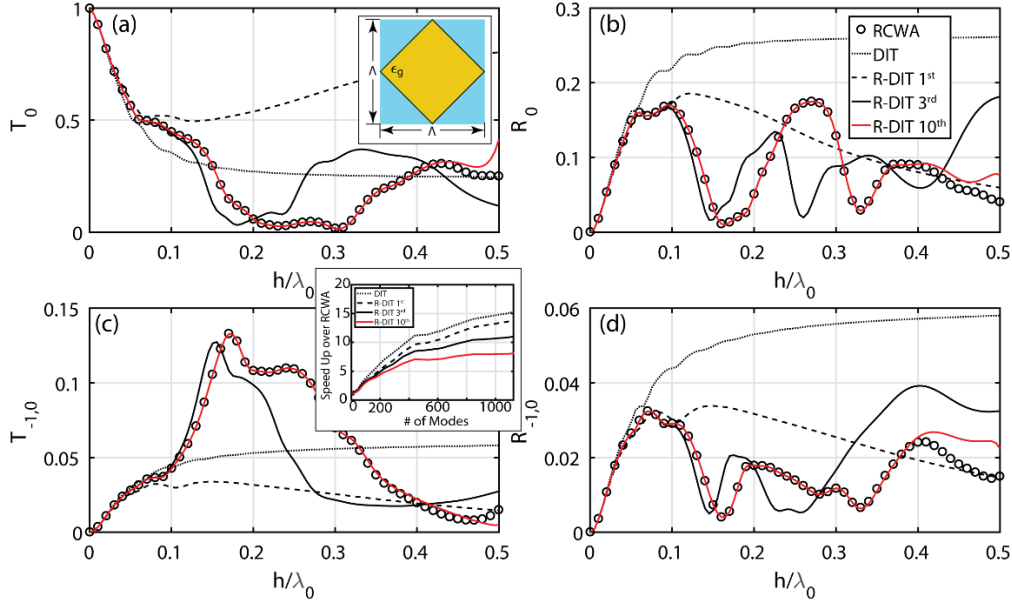

Figure 4: Specular and 1st order Reflection and Transmission results for RCWA(open circles), DIT(dottend line), and multiple different orders of R-DIT(dashed black line 1st order, solid black line 3rd order, and solid red line 10th order) for a free standing checkerboard ($\Lambda = 11.26 \mu m$) pattern composed of a lossy dielectric ($\epsilon = 10 + 1i$) in vacuum. (a) Specular Transmission, (b) Specular Reflection, (c) 1st Order diffracted transmission, and (d) 1st Order diffracted reflection. (inset c) Computational speed up of (R-)DIT over RCWA with (GP)GPU computing.

To conclude, we have presented a rigorous expansion of the generalized metasurface boundary condition which allows fast, accurate solutions of the optics of metasurfaces. We have shown agreement between the proposed R-DIT expansion with full-wave calculations for a variety of structures. R-DIT allows the calculation of optical properties of a metasurface, predicting both the direction and amplitude of diffracted waves with high accuracy, paving the way for rapid design of the next generation of metasurfaces.

This research has been supported by the NSF (grant DMR-#1209761).

**Appendix A. Illustration of matrices involved in construction of RCWA and R-DIT**:

For an isotropic homogeneous material, $\partial_x \equiv i k_x, \partial_y \equiv i k_y$, leading to

$$\widehat{\mathbb{P}} = \frac{c^2}{\omega^2}\begin{bmatrix} \frac{k_x k_y}{\epsilon_g} & \frac{\mu_g \omega^2}{c^2} - \frac{k_x^2}{\epsilon_g} \\ \frac{k_y^2}{\epsilon_g} - \frac{\mu_g \omega^2}{c^2} & -\frac{k_x k_y}{\epsilon_g} \end{bmatrix}; \; \widehat{\mathbb{Q}} = \frac{c^2}{\omega^2}\begin{bmatrix} -\frac{k_x k_y}{\mu_g} & \frac{k_x^2}{\mu_g} - \frac{\epsilon_g \omega^2}{c^2} \\ \frac{\epsilon_g \omega^2}{c^2} - \frac{k_x^2}{\mu_g} & \frac{k_x k_y}{\mu_g} \end{bmatrix}.$$

Consequently, the $\widehat{\mathcal{M}}$ matrix takes on a simple form:

$$\widehat{\mathcal{M}} \equiv \begin{bmatrix} 0 & 0 & \frac{k_x k_y}{\epsilon_g} & \frac{\mu_g \omega^2}{c^2} - \frac{k_x^2}{\epsilon_g} \\ 0 & 0 & \frac{k_y^2}{\epsilon_g} - \frac{\mu_g \omega^2}{c^2} & -\frac{k_x k_y}{\epsilon_g} \\ -\frac{k_x k_y}{\mu_g} & \frac{k_x^2}{\mu_g} - \frac{\epsilon_g \omega^2}{c^2} & 0 & 0 \\ \frac{\epsilon_g \omega^2}{c^2} - \frac{k_x^2}{\mu_g} & \frac{k_x k_y}{\mu_g} & 0 & 0 \end{bmatrix} \quad (A.1)$$

reducing Eq.(2) to an eigenvalue problem

$$\det\left(\frac{c}{\omega}\widehat{\mathcal{M}} - k_z \hat{I}\right) = \left(k_x^2 + k_y^2 + k_z^2 - \epsilon\mu\frac{\omega^2}{c^2}\right)^2 = 0 \quad (A.2)$$

that describes the dispersion of modes (plane waves) propagating through the media. Overall, Eq.(A.2) describes dispersion of four waves, that in case of isotropic homogeneous material are degenerate with respect to polarization (TE/TM) and their propagation direction ($\pm k_z$). Substitution of Eq.(A.2) into Eq.(2) yields a set of four eigen waves, whose in-plane components are given by

$$\begin{bmatrix}\mathcal{H}_x^{TE}\\ \mathcal{H}_y^{TE}\\ \mathcal{E}_x^{TE}\\ \mathcal{E}_y^{TE}\end{bmatrix} = \begin{bmatrix}-\frac{k_y c}{\omega}\\ \frac{k_x c}{\omega}\\ \mp\frac{k_x k_z c^2}{\mu\omega^2}\\ \mp\frac{k_x k_z c^2}{\mu\omega^2}\end{bmatrix},\quad \begin{bmatrix}\mathcal{H}_x^{TM}\\ \mathcal{H}_y^{TM}\\ \mathcal{E}_x^{TM}\\ \mathcal{E}_y^{TM}\end{bmatrix} = \begin{bmatrix}\frac{k_x c}{\omega}\\ \frac{k_y c}{\omega}\\ \mp\frac{\epsilon k_y}{k_z}\\ \pm\frac{\epsilon k_y}{k_z}\end{bmatrix}. \quad (A.3)$$

Where the top(bottom) signs correspond to waves propagating in the $+z$ ($-z$) direction.

Combining these solutions together, the field matrix $\widehat{\mathbb{F}^{(l)}}(z)$ [Eq.(7)] is given by

$$\widehat{\mathbb{F}^{(l)}}(z) \equiv \begin{bmatrix}-\frac{k_y c}{\omega} & \frac{k_x c}{\omega} & -\frac{k_y c}{\omega} & \frac{k_x c}{\omega}\\ \frac{k_x c}{\omega} & \frac{k_y c}{\omega} & \frac{k_x c}{\omega} & \frac{k_y c}{\omega}\\ -\frac{k_x k_z^{(l)} c^2}{\mu^{(l)}\omega^2} & -\frac{\epsilon^{(l)} k_y}{k_z^{(l)}} & \frac{k_x k_z^{(l)} c^2}{\mu^{(l)}\omega^2} & \frac{\epsilon^{(l)} k_y}{k_z^{(l)}}\\ -\frac{k_y k_z^{(l)} c^2}{\mu^{(l)}\omega^2} & \frac{\epsilon^{(l)} k_x}{k_z^{(l)}} & \frac{k_y k_z^{(l)} c^2}{\mu^{(l)}\omega^2} & -\frac{\epsilon^{(l)} k_x}{k_z^{(l)}}\end{bmatrix} \times \begin{bmatrix}e^{ik_z^{(l)} z} & 0 & 0 & 0\\ 0 & e^{ik_z^{(l)} z} & 0 & 0\\ 0 & 0 & e^{-ik_z^{(l)} z} & 0\\ 0 & 0 & 0 & e^{-ik_z^{(l)} z}\end{bmatrix}$$

[1] N. K. Grady, J. E. Heyes, D. R. Chowdhury, Y. Zeng, M. T. Reiten, A. K. Azad, A. J. Taylor, D. A. Dalvit, and H.-T. Chen. Terahertz metamaterials for linear polarization conversion and anomalous refraction. Science, 340(6138):1304–1307, 2013.

2 Z. Bomzon and E. Hasman. The formation of laser beams with pure azimuthal or radial polarization. Appl. Phys. Lett., 77(21), 2000.

3 P. Cencillo-Abad, E. Plum, E. T. F. Rogers, and N. I. Zheludev. Spatial optical phase-modulating metadevice with subwavelength pixelation. Opt. Express, 24(16):18790–18798, Aug 2016.

4 N. Yu, P. Genevet, M. A. Kats, F. Aieta, J.-P. Tetienne, F. Capasso, and Z. Gaburro. Light propagation with phase discontinuities: generalized laws of reflection and refraction. Science, 334(6054):333–337, 2011.

5 S. Sun, Q. He, S. Xiao, Q. Xu, X. Li, and L. Zhou. Gradient-index meta-surfaces as a bridge linking propagating waves and surface waves. Nat. Materials, 11(5):426–431, 2012.

6 X. Ni, N. K. Emani, A. V. Kildishev, A. Boltasseva, and V. M. Shalaev. Broadband light bending with plasmonic nanoantennas. Science, 335(6067):427–427, 2012.

7 Y. Zhao and A. Alù. Manipulating light polarization with ultrathin plasmonic metasurfaces. Phys. Rev. B, 84:205428, Nov 2011.

8 A. V. Kildishev, A. Boltasseva, and V. M. Shalaev. Planar photonics with metasurfaces. Science, 339(6125), 2013.

9 Y. Liu and X. Zhang. Metasurfaces for manipulating surface plasmons. Appl. Phys. Lett., 103(14):–, 2013.

10 P.-Y. Chen and A. Alù. Mantle cloaking using thin patterned metasurfaces. Phys. Rev. B, 84:205110, Nov 2011.

11 L. Huang, X. Chen, H. Mühlenbernd, H. Zhang, S. Chen, B. Bai, Q. Tan, G. Jin, K.-W. Cheah, C.-W. Qiu, et al. Three-dimensional optical holography using a plasmonic metasurface. Nature communications, 4, 2013.
12 X. Ni, A. V. Kildishev, and V. M. Shalaev. Metasurface holograms for visible light. Nature communica- tions, 4, 2013.
13 G. Zheng, H. Mühlenbernd, M. Kenney, G. Li, T. Zentgraf, and S. Zhang. Metasurface holograms reaching 80% efficiency. Nature nanotechnology, 10(4):308–312, 2015.
14 C. M. Roberts, S. Inampudi, and V. A. Podolskiy. Diffractive interface theory: nonlocal susceptibility approach to the optics of metasurfaces. Opt. Express, 23(3):2764–2776, Feb 2015.
15 C. L. Holloway, A. Dienstfrey, E. F. Kuester, J. F. O'Hara, A. K. Azad, and A. J. Taylor. A discussion on the interpretation and characterization of metafilms/metasurfaces: The two-dimensional equivalent of metamaterials. Metamaterials, 3(2):100 – 112, 2009.
16 A. Epstein and G. V. Eleftheriades. Huygens' metasurfaces via the equivalence principle: design and applications. J. Opt. Soc. Am. B, 33(2):A31–A50, Feb 2016.
17 G. Eleftheriades and A. Alù. Electromagnetic metasurfaces: introduction. J. Opt. Soc. Am. B, 33(2):EM1–EM1, Feb 2016.
18 D. Zaluški, A. Grbic, and S. Hrabar. Analytical and experimental characterization of metasurfaces with normal polarizability. Phys. Rev. B, 93:155156, Apr 2016.
19 S. A. H. Nekuee, A. Khavasi, and M. Akbari. Highly accurate and fast convergent diffractive interface theory for fast analysis of metasurfaces. IEEE Journal of Quantum Electronics, 52(7):1–6, 2016.
20 J. Jin. Finite Element-Boundary Element Methods for Electromagnetic Scattering. University of Michi- gan, 1989.
21 K. S. Kunz and R. J. Luebbers. The finite difference time domain method for electromagnetics. CRC press, 1993.
22 M. G. Moharam and T. K. Gaylord. Rigorous coupled-wave analysis of planar-grating diffraction. J. Opt. Soc. Am., 71(7):811–818, Jul 1981.
23 M. G. Moharam, T. K. Gaylord, E. B. Grann, and D. A. Pommet. Formulation for stable and efficient implementation of the rigorous coupled-wave analysis of binary gratings. J. Opt. Soc. Am. A, 12(5):1068–1076, May 1995.
24 L. Li. New formulation of the fourier modal method for crossed surface-relief gratings. JOSA A, 14(10):2758–2767, 1997.
25 P. Yeh, A. Yariv, and C.-S. Hong. Electromagnetic propagation in periodic stratified media. i. general theory. J. Opt. Soc. Am., 67(4):423–438, Apr 1977.
26 T. Schuster, J. Ruoff, N. Kerwien, S. Rafler, and W. Osten. Normal vector method for convergence improvement using the rcwa for crossed gratings. JOSA A, 24(9):2880–2890, 2007.
27 J. Chandezon, M. T. Dupuis, G. Cornet, and D. Maystre. Multicoated gratings: a differential formalism applicable in the entire optical region. J. Opt. Soc. Am., 72(7):839–846, Jul 1982.
28 M. Nevière and E. Popov. Light propagation in periodic media: differential theory and design. CRC Press, 2002.
29 S. Law, D. C. Adams, A. M. Taylor, and D. Wasserman. Mid-infrared designer metals. Opt. Express, 20(11):12155–12165, May 2012.
30 http://ark.intel.com/products/40201/Intel-Xeon-Processor-L5520-8M-Cache-2_26-GHz-5_86-GTs-Intel-QPI (Retrieved: Sept. 2016)
31 NVIDA. Kepler compute architecture white paper. Technical report, NVIDIA Corperation, 2012.
32 J. W. Demmel, J. Dongarra, B. Parlett, W. Kahan, M. Gu, D. Bindel, Y. Hida, X. Li, O. Marques, E. J. Riedy, C. Vömel, J. Langou, P. Luszczek, J. Kurzak, A. Buttari, J. Langou, and S. Tomov. Prospectus for the Next LAPACK and ScaLAPACK Libraries, pages 11–23. Springer Berlin Heidelberg, Berlin, Heidelberg, 2007.
33 COMSOL Multiphysics, http://www.comsol.com (Retrieved: Sept. 2016)